\documentclass[twocolumn,showpacs,preprintnumbers,amsmath,amssymb]{revtex4}

\usepackage{graphicx}
\usepackage{dcolumn}
\usepackage{bm}

\begin{document}

\preprint{APS/123-QED}

\title{Storage by trapping and spatial staggering \\of multiple interacting solitons in $\Lambda$-type media}

\author{Willem P. Beeker$^{1}$, Chris J. Lee$^{1}$, Edip Can$^{2}$, and Klaus -J. Boller$^{1}$}\email{k.j.boller@tnw.utwente.nl}
\homepage{http://lpno.tnw.utwente.nl/}
 \affiliation{$^{1}$Laser Physics \& Nonlinear Optics Group, MESA+ Research Institute for Nanotechnology, 
University of Twente, P. O. Box 217, Enschede 7500AE, The Netherlands 
\\
 $^{2}$Physics of Fluids Group, Faculty of Science and Technology, University of Twente, The Netherlands
}

\date{\today}

\begin{abstract}
In this paper we investigate the properties of self induced transparency (SIT) solitons, propagating in a $\Lambda$-type medium. We find that the interaction between SIT solitons can lead to trapping with their phase preserved in the ground state coherence of the medium. These phases can be altered in a systematic way by the application of appropriate light fields, such as additional SIT solitons. Furthermore, multiple independent SIT solitons can be made to propagate as bi-solitons through their mutual interaction with a separate light field. Finally, we demonstrate that control of the SIT soliton phase can be used to implement an optical exclusive-or gate.
\end{abstract}

\pacs{42.65.Tg, 42.79.Ta, 42.50.Gy}
\maketitle

\section{introduction}
\label{s:introduction}
Strong and coherent interactions between $\Lambda$-type three-level atomic media and resonant light pulses have been a topic of much interest recently. Driving this interest are various optical communications applications, such as optical buffers and phase shifters for optical routers~\cite{Heebner, Melloni}. Of particular interest is all-optical switching based on solitons to realize optical gates  (\cite{Scheuer} and references therein). Generally, the Kerr effect is chosen as the nonlinear interaction between solitons due to its presence in many different media. However, the low value of typical Kerr coefficients implies that intense optical pulses and/or long interaction lengths must be used. In contrast, self-induced transparency~\cite{mccall} (SIT) provides solitons that are based on a strong, resonant optical nonlinear effect, which could, in principle, reduce interaction lengths and pulse intensities required for all-optical switching.

\begin{figure}
\includegraphics[width=60mm]{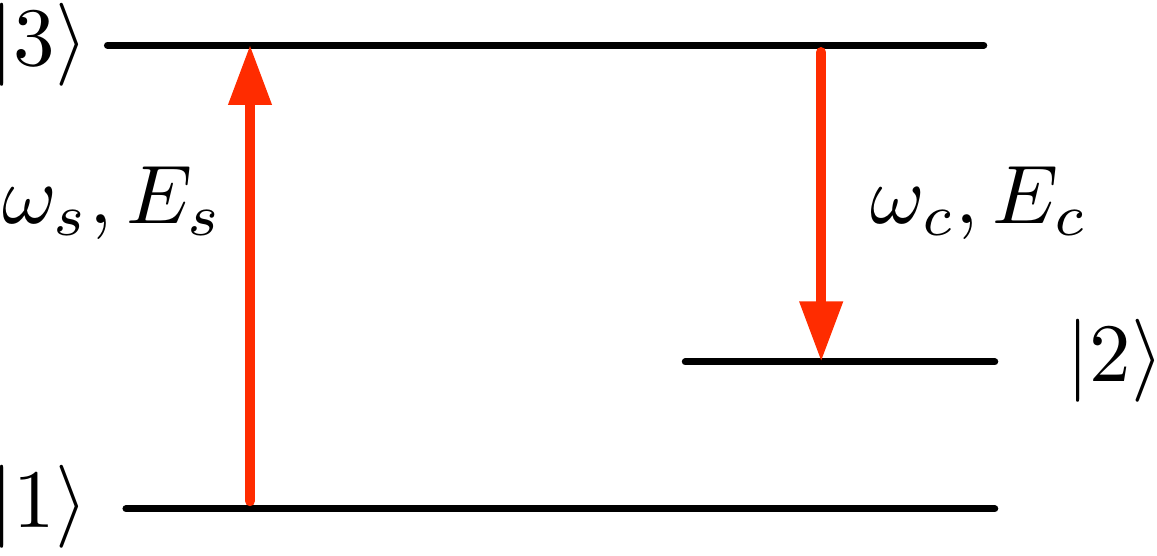}
\caption{\label{fig:levelscheme}$\Lambda$-three level system with two ground states and one excited state. A transition from state $\left|1\right\rangle$  to $\left|3\right\rangle$  can be performed with a SIT soliton. The transition from $\left|2\right\rangle$  to $\left|3\right\rangle$  can be performed with control light field. A radiative transition between $\left|1\right\rangle$  and $\left|2\right\rangle$  is dipole forbidden.}
\end{figure}

The interaction of a single SIT soliton with a control pulse has been investigated by Vemuri \textit{et al.} \cite{vemuri}. They showed that a single SIT soliton can be cloned in a $\Lambda$-type medium by applying a weak control light field resonant with the $\left|2\right\rangle$-$\left|3\right\rangle$ transition (see Fig.~\ref{fig:levelscheme}) .  In doing so, the spatial region of the medium where the SIT soliton and control pulse overlap is left with a significant population in $\left|2\right\rangle$ and a ground state coherence extending to either side of the region \cite{loiko2006, loiko2007}, which we refer to as the trapped soliton. 

The trapped soliton presents an intriguing possibility: namely, that linear and nonlinear optical operations may be performed on the trapped soliton using resonant light fields. Normally, operations between different solitons must be carefully timed so that the overlap in the linear (e.g., beam-splitters) or nonlinear optical elements is maximized. However, in the case of a trapped soliton, the timing problem is much less critical because optical operations can be performed on the stationary coherence, rather than on the traveling soliton. 

Unfortunately, very little is known about the quantum or classical interactions between trapped solitons, propagating SIT solitons, or other light fields. Recent work has shown that the presence of a captured soliton is sufficient to trap additional solitons \cite{loiko2006,Beeker}, however, the mechanism of such trapping, and the interaction of the trapped solitons with light fields resonant with the $|2\rangle$-$|3\rangle$ transition remained unexplored. In what follows, we go beyond these results by examining the influence of additional control pulses and SIT solitons on the state of the medium and the phase of the emitted light. 

The details of the trapping process are explained and we highlight this by examining three representative cases. The interaction of a control light field with both single captured solitons and multiple solitons is examined. We found that multiple trapped solitons can behave like bi-solitons. Further, we show that the trapping process and interactions with control light fields manipulates the phases of the captured soliton. Finally, by choosing the soliton phase as an information carrier, we demonstrate a series of operations that implement an all-optical XOR gate.

\section{theory}
\label{s:theory}
To study the propagation and interaction of light in a three-level medium, where each atom has energy states as shown in Fig.~\ref{fig:levelscheme}, we employ a density matrix approach in the rotating wave approximation (RWA)~\cite{milonni}. The medium's constituent atoms have one excited state and two ground states. The $\left|1\right\rangle$-$\left|3\right\rangle$  and $\left|2\right\rangle$-$\left|3\right\rangle$ are dipole allowed radiative transitions, while the $\left|1\right\rangle$-$\left|2\right\rangle$ transition is dipole forbidden. The medium is discretized along the propagation axis, creating a 1-D grid. At each grid location along the propagation axis, the medium's state is represented by an independent density matrix. This leads to the following set of differential equations for each grid point, the solution of which provides the time and space-dependent population densities and coherences between populations throughout the whole medium. 
\begin{eqnarray}
\label{eq:rho11}
\dot{\rho}_{11}&=-\frac{i}{2}\left(\chi_{13}\rho_{13}-\chi^{*}_{13}\rho^{*}_{13}\right)+\rho_{33}R_{31}\\
\label{eq:rho22}
\dot{\rho}_{22}&=-\frac{i}{2}\left(\chi_{23}\rho_{23}-\chi^{*}_{23}\rho^{*}_{23}\right)+\rho_{33}R_{32}\\
\dot{\rho}_{33}&=\frac{i}{2}\left(\chi_{13}\rho_{13}+\chi_{23}\rho_{23}-\chi^{*}_{13}\rho^{*}_{13}-\chi^{*}_{23}\rho^{*}_{23}\right)\nonumber\\
\label{eq:rho33}
&-\left(R_{31}+R_{32}\right)\rho_{33}\\
\label{eq:rho12}
\dot{\rho}_{12}&=\frac{i}{2}\left(\chi^{*}_{13}\rho^{*}_{23}-\chi_{23}\rho_{13}\right)\\
\label{eq:rho13}
\dot{\rho}_{13}&=\frac{i}{2}\left(\chi^{*}_{13}\left(\rho_{33}-\rho_{11}\right)-\chi^{*}_{23}\rho_{12}\right)-\beta_{13}\rho_{13}\\
\label{eq:rho23}
\dot{\rho}_{23}&=\frac{i}{2}\left(\chi^{*}_{23}\left(\rho_{33}-\rho_{22}\right)-\chi^{*}_{13}\rho^{*}_{12}\right)-\beta_{23}\rho_{23}
\end{eqnarray}

The diagonal elements of the density matrix, $\rho_{ii}$, are the population densities, normalized such that Tr($\rho$) = 1, of the energetic states of the medium. The off-diagonal elements, $\rho_{ij}$, are the coherence terms that are proportional to the polarization of the medium. The decay constants $R_{31}$ and $R_{32}$ account for spontaneous emission from $\left|3\right\rangle$ to $\left|1\right\rangle$ and $\left|2\right\rangle$, respectively. Decay of the coherence between states $i$ and $j$ is given by the rate, $\beta_{ij}$, representing non-radiative processes, such as collisional dephasing. The relationship between the Rabi frequencies, $\chi_{ij}$, and the applied light pulses field envelopes $E_s$ and $E_c$, is given by

\begin{eqnarray}
\label{eq:chi13}
\chi_{13}=\frac{E_{s}\mu_{13}}{\hbar}\\
\label{eq:chi23}
\chi_{23}=\frac{E_{c}\mu_{13}}{\hbar}
\end{eqnarray}
where $\mu_{13}$ and $\mu_{23}$ are the transition dipole moments of the two radiative transitions.

The light fields are assumed to be resonant with their respective transitions. The duration of the light pulses are assumed to be much shorter than the spontaneous emission rates, but long enough for the slowly varying envelope approximation (SVEA) to hold. We also require that the life-time of $\left|3\right\rangle$ is sufficient to allow stimulated emission processes dominate spontaneous emission. The temporal profile of the light field envelopes are given by sech-squared functions, as SIT solitons assume this form naturally \cite{mccall}.

Applying the SVEA and limiting ourselves to one spatial dimension (plane wave propagation into the positive z-direction), the Maxwell equations are given by:
\begin{eqnarray}
\label{eq:Es}
\frac{dE_{s}}{dz}=i\frac{k_{s}}{\epsilon_{0}}N\mu_{13}\rho^{*}_{13}\\
\label{eq:Ec}
\frac{dE_{c}}{dz}=i\frac{k_{c}}{\epsilon_{0}}N\mu_{23}\rho^{*}_{23}
\end{eqnarray}
with $k_{s}$ and $k_{c}$ the wavenumbers of the SIT and coupling fields, $N$ the density of the active medium and $\epsilon_{0}$ the electric permittivity of the vacuum.

The coupled set of density and Maxwell equations Eqs.~\ref{eq:rho11}-\ref{eq:rho23}, \ref{eq:Es} and \ref{eq:Ec}, were solved numerically using a fourth order Runge-Kutta algorithm. All calculations begin with the $\left|1\right\rangle$ state fully occupied ($\rho_{11}$ = 1) and no coherence between any of the levels.

In our calculations we use a value for the product of $N\mu=1.7\cdot10^{-10}$~Cm$^{-2}$ \cite{hilborn}, which corresponds, for example, to probing the transition from state 4$^1$P to state 3$^1$D of calcium in a gas-cell with a 30~mBar partial pressure of calcium. 
 
After a number of solitons and control pulses have interacted with the medium, we examine the spatial structure of the ground state coherence ($\rho_{12}$), the phase of the ground state coherence, and the phase of light emitted by the medium. In general the phase of the light envelope is given by the argument of $E_{i} = A_{i}exp(i\phi_{i})$. Likewise the phase of the ground state coherence is given by the argument of $\rho_{12} = A_{\rho} exp(i\theta)$. The phases are defined with respect to an initial control pulse, which we consider to have a phase of zero. In other words, the control pulse has a purely real field envelope.

\section{results}
\label{s:results}
\begin{figure}
\includegraphics[width=\columnwidth]{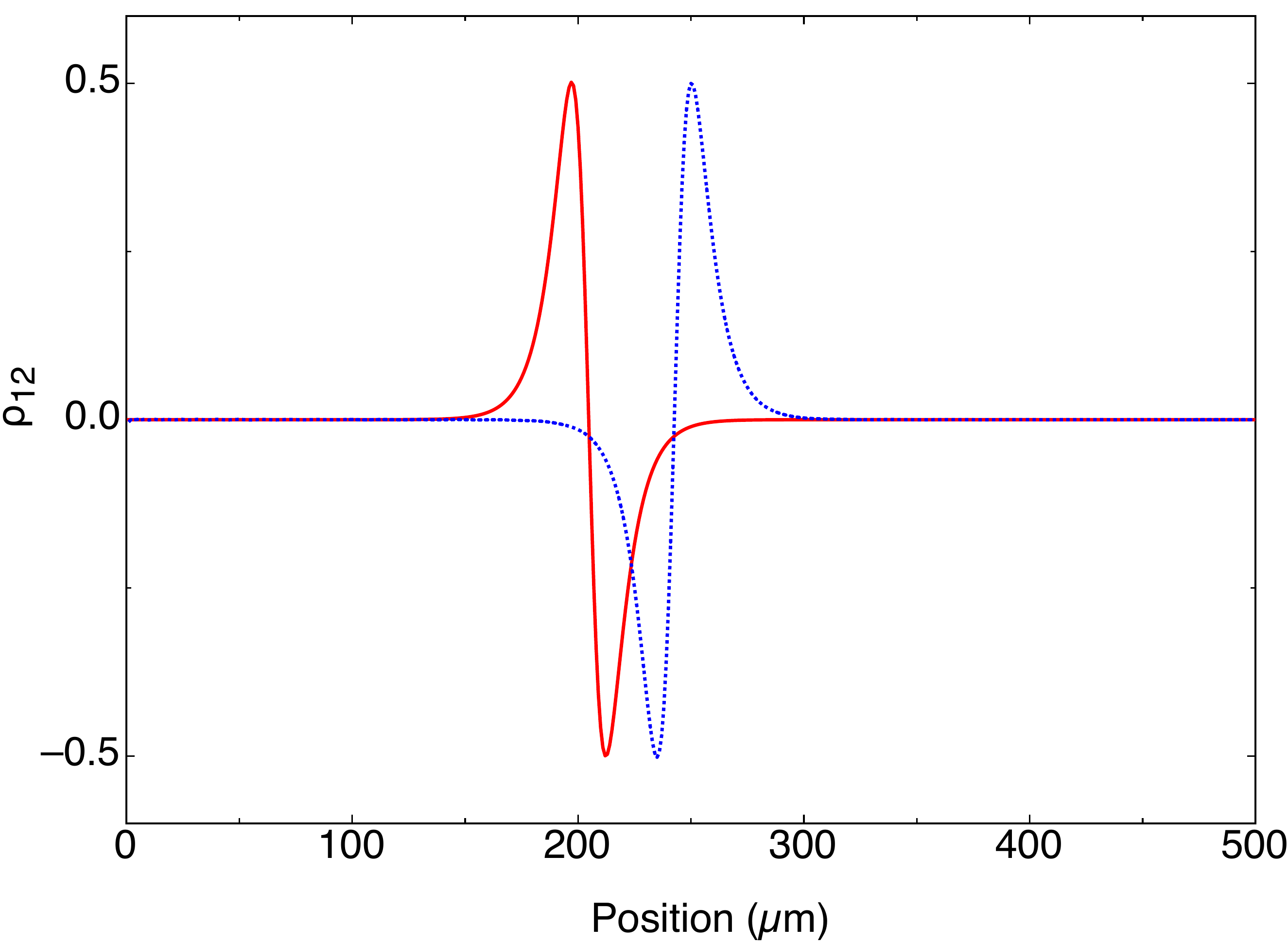}
\caption{\label{fig:pulseafterpi}State of the ground state coherence after the storage of a single SIT soliton by a small coupling pulse (red trace). Ground state coherence after the application of a sufficiently intense control pulse to a medium containing a captured SIT soliton (blue dotted trace). The excited region is shifted deeper into the medium and the ground state coherence undergoes a $\pi$ phase shift.}
\end{figure}

We begin by partially repeating the work of Vemuri et al. \cite{vemuri}, where a single SIT soliton is trapped by the application of a single control pulse. This occurs because as the SIT soliton excites population into $\left|3\right\rangle$, the control pulse stimulates emission into $\left|2\right\rangle$. The result is that the control pulse is amplified at the expense of the SIT soliton, which is absorbed by the medium. As shown by Loiko et al. \cite{loiko2007}, in this process, the medium has stored the sech-squared shape of the SIT soliton in the medium as a sech-squared shaped distribution of population in the second ground state. As a first phenomenon we observe that, in addition to the shape preservation, the phase of the stored soliton is also stored in the ground state coherence of the medium (see red trace in Fig.~\ref{fig:pulseafterpi}). More precisely:
\begin{eqnarray}
\theta=\phi_{23}-\phi_{13}
\label{eq:eleven}
\end{eqnarray}
where $\phi_{23}$ is the phase of the control pulse and $\phi_{13}$ is the phase of the SIT soliton. This relationship comes from the fact that the coherence of the medium is proportional to the polarization of the medium, which opposes the light fields. Our calculations show another phenomenon: The blue dotted trace in Fig.~\ref{fig:pulseafterpi} shows the resulting ground state coherence after a sufficiently intense control pulse has interacted with the captured SIT soliton. Note that this control pulse shifts the SIT soliton's phase by $\pi$, while also shifting the location of the captured SIT soliton. However, control pulses with a relatively low intensity shift the location of the ground state coherence without altering its phase~\cite{LoikoFootnote}.

\subsection{Interactions between SIT solitons}
\label{ss:interactions}
We now turn to the case where a second SIT soliton is injected into the medium after the capture of a first SIT soliton. In this case, the second SIT soliton encounters a medium with populations and coherences described above and illustrated in Fig.~\ref{fig:pulseafterpi}. As will be described in more detail, we observed that, in general, these additional SIT solitons are trapped as well upon ``collision'' with the excited region, and that this leads to the creation of additional excited regions. An important feature of this process is that the second and subsequent SIT solitons are trapped without a control pulse being injected.

\begin{figure} 
\includegraphics[width=\columnwidth]{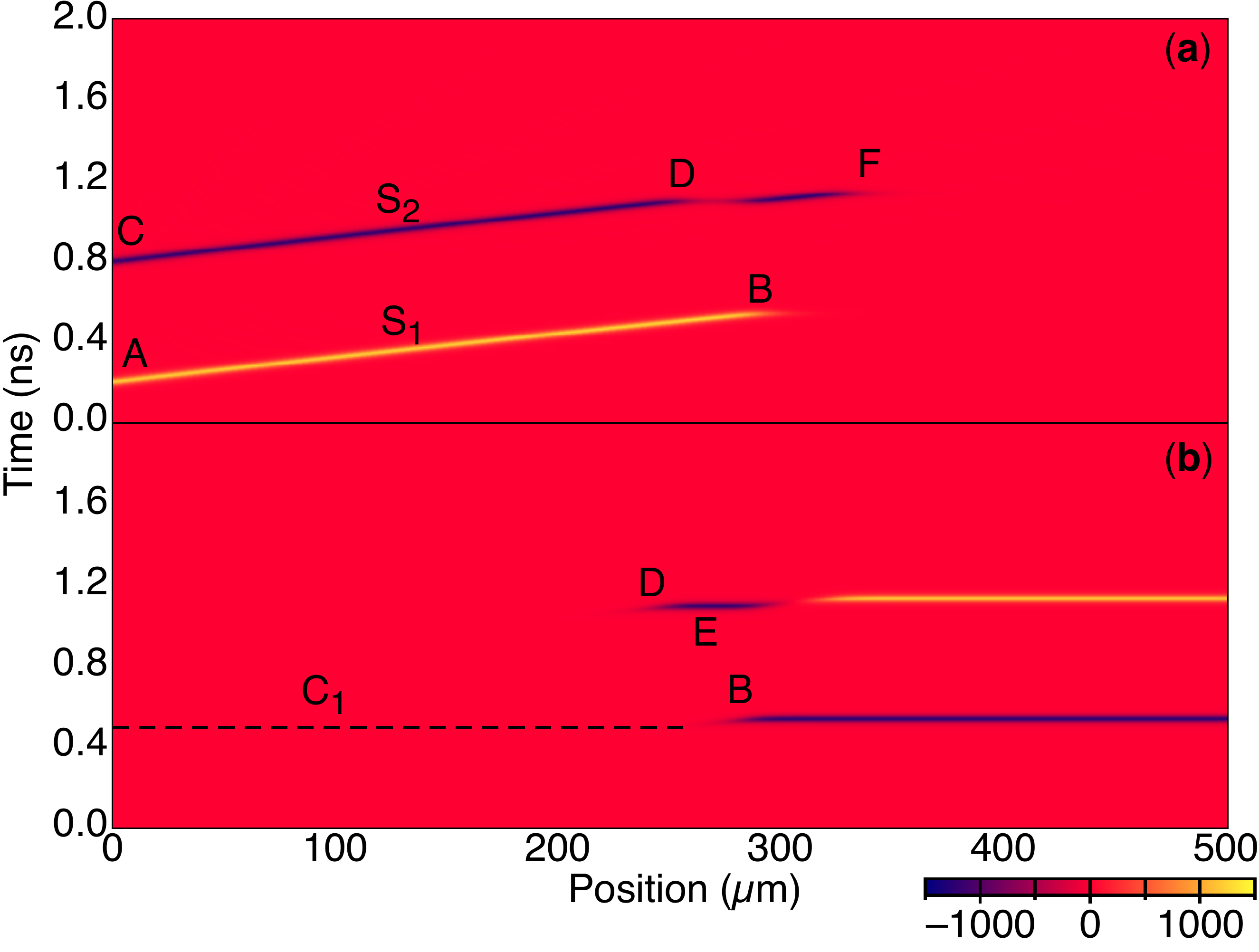}
\caption{\label{fig:lightfields}Electric field amplitude of light resonant with the $\left|1\right\rangle$-$\left|3\right\rangle$ transition, as either an injected SIT soliton, or emission from the medium. (a) Electric field amplitude (Vm$^{-1}$) of light resonant with the $|1\rangle$-$|3\rangle$ transition and (b) Electric field amplitude (Vm$^{-1}$) of light resonant with the $\left|2\right\rangle$-$\left|3\right\rangle$ transition, as either an injected control pulse, or emission from the medium. Event $A$: the first SIT soliton is introduced to the medium. Event $B$: the SIT soliton and control pulse overlap, resulting in the SIT soliton's capture and the coupling pulse's amplification. Event $C$: the second SIT soliton is introduced to the medium. Event $D$: the SIT soliton collides with the first half of the excited region, emitting a negative control pulse. Event $E$: the negative control pulse captures the SIT soliton and is amplified. It then destructively interferes with the ground state coherence of the second half of the excited region, resulting in the emission of a transitory SIT soliton. Event $F$: the remaining control light field is amplified by the transitory SIT soliton, capturing it in the process and extending the original excitation region.}
\end{figure}

This can be understood by examining the behavior of the SIT soliton and the $\Lambda$-system, illustrated in Fig.~\ref{fig:lightfields}. Fig.~\ref{fig:lightfields}(a) shows the intensity of the SIT solitons as a function of time and space, while Fig.~\ref{fig:lightfields}(b) shows the intensity of control light fields. An initial SIT soliton is injected into the medium at time $A$, which travels slowly through the medium. A short time later, a control pulse is injected, it is, however, too weak to be visible on the scale of Fig.~\ref{fig:lightfields}(b). The control pulse overtakes the SIT soliton at time/position $B$, where it is amplified at the expense of the SIT soliton, producing the excited region illustrated in Fig.~\ref{fig:pulseafterpi}. Note that the amplified control pulse exits the medium without further perturbation. 

A second SIT soliton is injected into the medium at time $C$, however, no control light fields are injected. Nevertheless, the SIT soliton is trapped in the medium by the collision with the excited region at time/position $D$. When the SIT pulse enters the first half of the excited region, the ground state coherence ensures that there is simultaneous stimulated emission on both the $\left|1\right\rangle$-$\left|3\right\rangle$ and $\left|2\right\rangle$-$\left|3\right\rangle$ transitions. However, for this case, the light on the control transition is out of phase with the original control pulse, indicated by a negative value in Fig.~\ref{fig:lightfields}(b) at location $D$. This control pulse is amplified at the expense of the SIT soliton, creating a new excited region in the medium. 

At location $E$, the control light field enters the second half of the excited region, where the sign of the coherence is the reverse to the first half. The population transfer due to the control pulse removes this coherence, generating a new SIT soliton, which we refer to as a transitory SIT soliton. In doing so, the control pulse is completely absorbed and a new control light field with the phase of the original control pulse is emitted by the residual population and ground state coherence. This control pulse captures the newly generated SIT pulse, being amplified in the process and exits the medium (location $F$). The second capture process generates another excitation region. As a result, the medium now has a differently shaped excitation region, which is wider than before.

It is also important to note that the soliton collision results in a control pulse exiting the medium. This is a general result, and also occurs if collisions occur by inserting subsequent SIT pulses.

As described above, a transitory SIT soliton is emitted during the collision and subsequently recaptured by control light fields that are also emitted during the collision process. Thus, one would expect that the coherence generated during the capture of the transitory SIT soliton replicates that of the second SIT soliton capture. However, the original coherence, which is the source for all the control light fields after the first SIT soliton is captured, is an odd function with respect to the center location of the captured SIT soliton. Thus, the control light fields, emitted from each side of this location during the collision, are out of phase with each other. Consequently, these two light fields destructively interfere with each other. In this case, the two SIT solitons are identical, leading to a near complete destructive interference of the control light field. The control light that subsequently captures the transitory SIT soliton is generated from the residual excited population and coherence, which has a phase identical to the original control pulse, thus, the coherence associated with the transitory SIT soliton is identical to that of the first SIT soliton.

This is highlighted by examining three cases: $P_{pr} = P_{tr}$, $P_{pr} > P_{tr}$, $P_{pr} < P_{tr}$, where $P_{tr}$ is the peak intensity of the SIT soliton that is captured by a control light field and $P_{pr}$ is the peak intensity of the SIT soliton that collides with the ground state coherence, which was written into the medium due to the capture of the first SIT pulse. 

\textbf{Case 1: $P_{pr} = P_{tr}$}

\begin{figure} 
\includegraphics[width=\columnwidth]{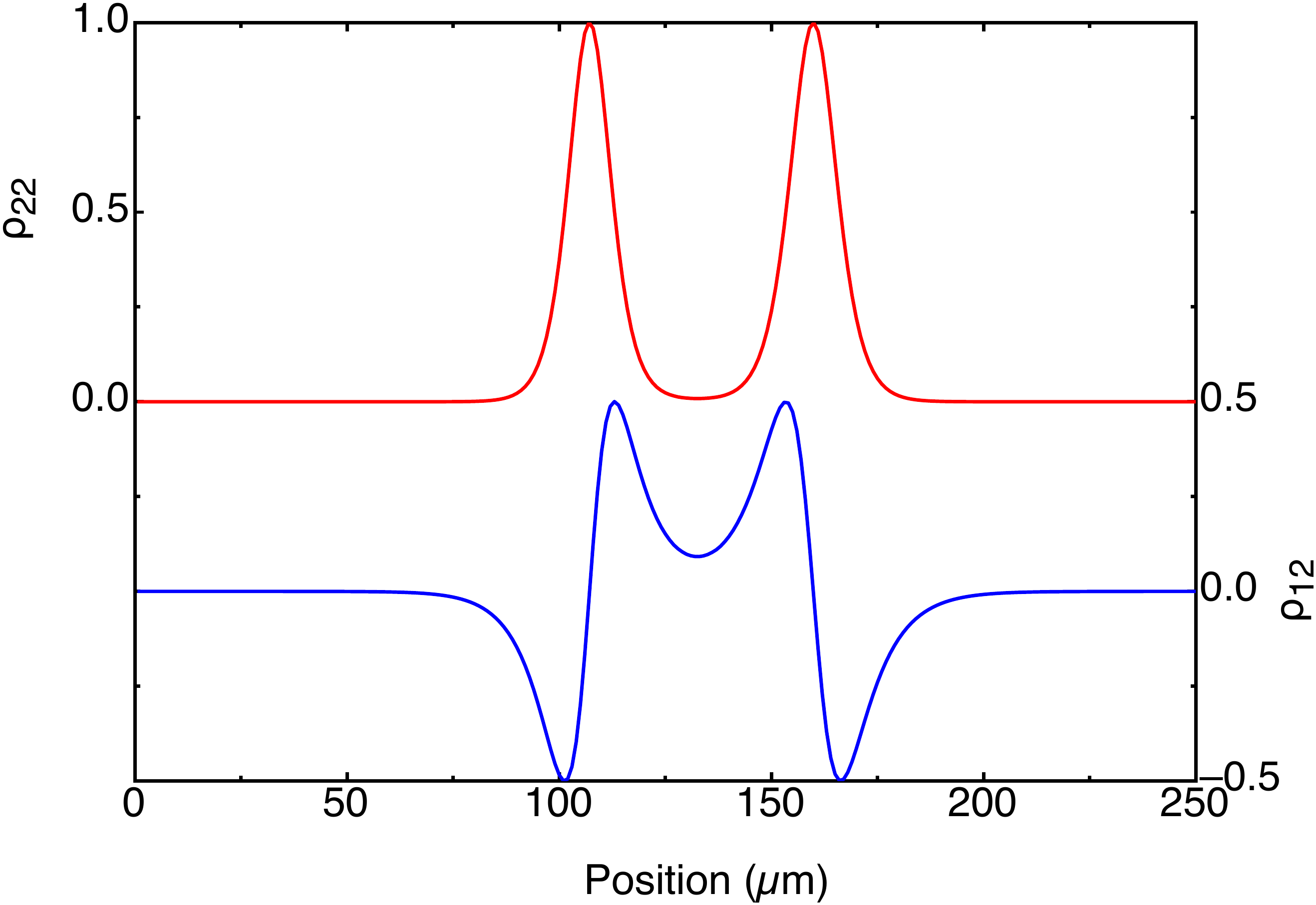}
\caption{\label{fig:Peq}$\rho_{22}$ (red trace) and ground state coherence (blue trace) through out the medium after two SIT solitons with equal peak intensity are captured.}
\end{figure}

For the case of equal intensity SIT solitons, the second SIT soliton collides with the excited region and, as a result, does not penetrate as far into the medium as the first SIT soliton (see Fig.~\ref{fig:Peq}) in comparison to Fig.~\ref{fig:pulseafterpi}. The act of colliding emits a transitory SIT soliton, which expands the excited region deeper into the medium by approximately twice its original full width half maximum (FWHM) width. The two peaks in $\rho_{22}$ are separated by 4 times the FWHM width of the original excited state regions. Note that $\rho_{22}$ falls to zero between the two peaks, however, as we will show later, these two regions remain coupled. 

Note also that the phase of the coherence appears to be inverted with respect to the original shape. The inversion occurs because the control light field that captures the second SIT soliton is generated by the coherence on the left hand side of the excited region in Fig.~\ref{fig:pulseafterpi}. This light is out of phase with respect to the original control pulse (we also refer to this as a negative control pulse), thus, from eqn.~\ref{eq:eleven}, a shift of $\pi$ is expected. 

\textbf{Case 2: $P_{pr} < P_{tr}$}

\begin{figure}
\includegraphics[width=\columnwidth]{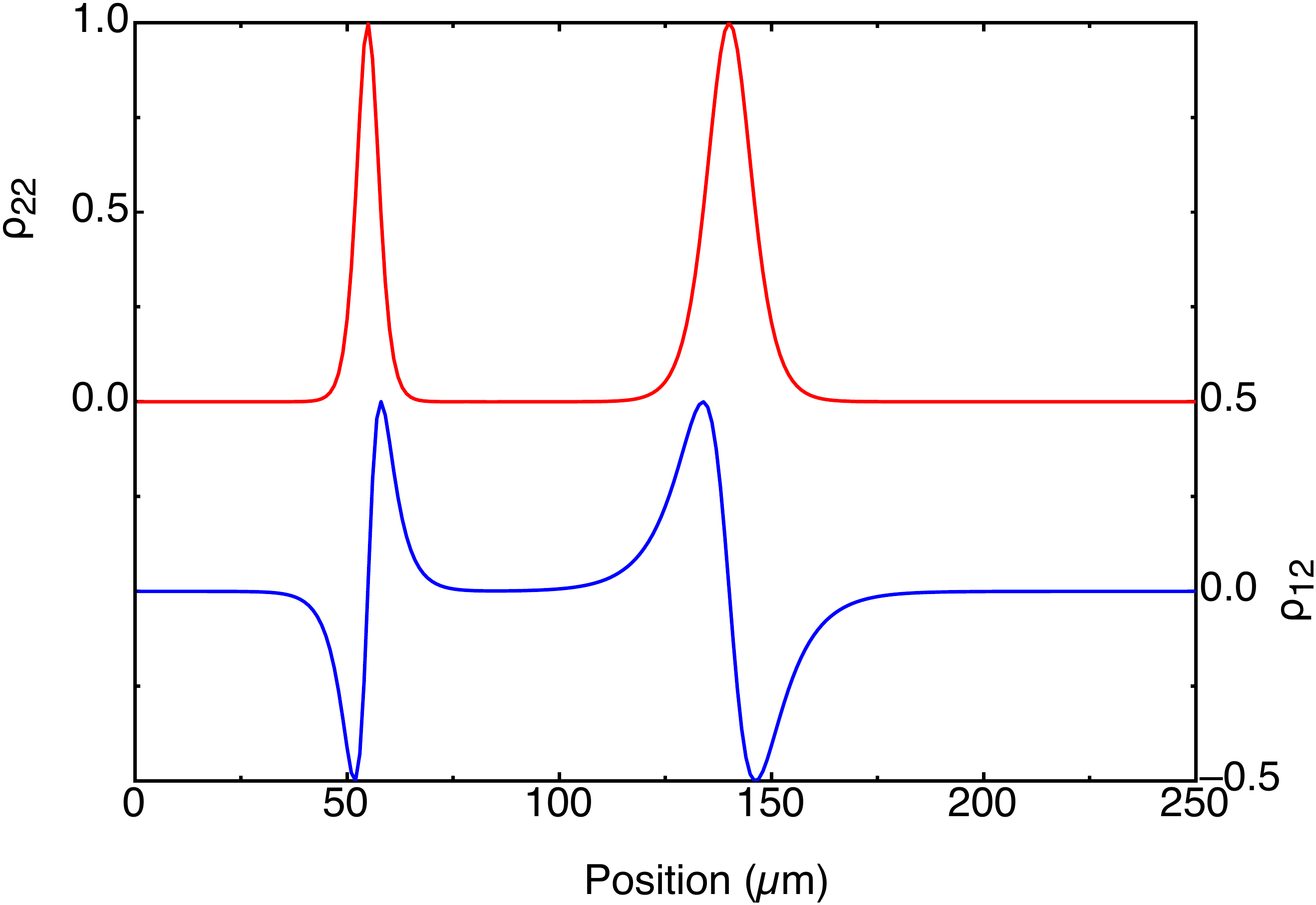}
\caption{\label{fig:PrSmaller}$\rho_{22}$ (red trace) and ground state coherence (blue trace) through out the medium after two SIT solitons are captured. In this case, the peak intensity of the second SIT soliton is less than that of the first.}
\end{figure}

As in the previous case and shown in Fig.~\ref{fig:PrSmaller}, the second SIT soliton is indeed trapped, however, the resulting state of the medium is somewhat different. The lower amplitude and slower traveling second SIT soliton emits a lower amplitude and slower traveling transitory SIT soliton during the collision. Consequently, the second $\rho_{22}$ peak is not shifted as deeply into the medium (60\% of its FWHM width) compared to case 1. The distance between both peaks is about 6.5 times the FWHM of the original excited region, which is greater than for case 1. The greater separation is due to the fact that the second injected SIT soliton travels slower and is thus trapped over a shorter distance by the emitted control light fields. 

As before, the excited population is surrounded by ground state coherence with the ground state coherence undergoing a $\pi$ phase transition when $\rho_{22}$ returns to zero between the two peaks. This indicates that, as in the previous case, the transitory SIT soliton is captured by a positive coupling pulse. Given case 1, where equal amplitude SIT solitons result in a positive coupling pulse, this is expected. The relatively less intense control fields, emitted during the trapping of the second SIT soliton on the left hand side of the original capture location, destructively interfere with the more intense control light fields emitted from the right hand side of the capture location. The end result is a control light field with the same phase as the original control pulse.

\textbf{Case 3: $P_{pr} > P_{tr}$}

\begin{figure}
\includegraphics[width=\columnwidth]{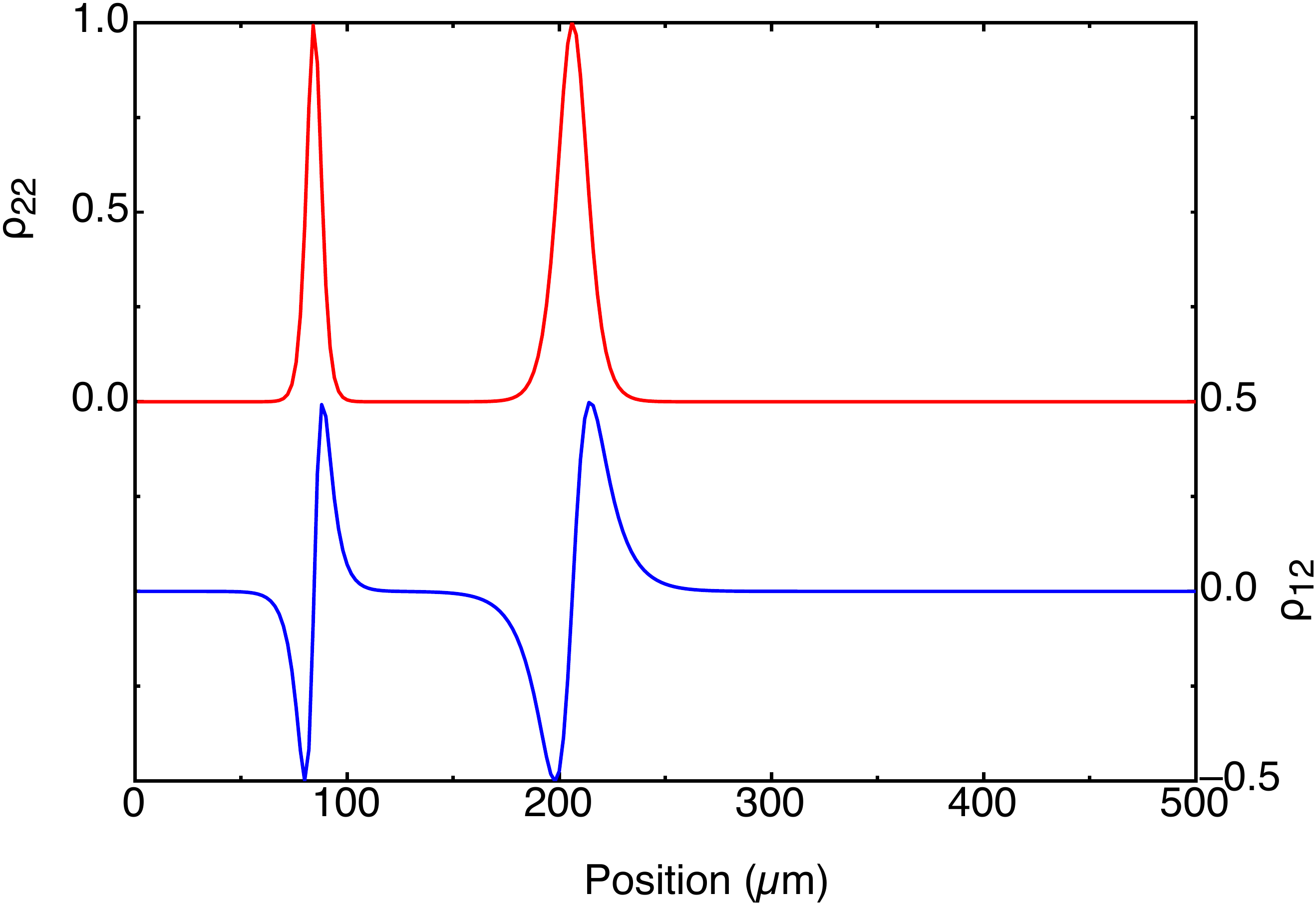}
\caption{\label{fig:PrBigger}$\rho_{22}$ (red trace) and ground state coherence (blue trace) through out the medium after two SIT solitons are captured. In this case, the peak intensity of the first SIT soliton is less than that of the second.}
\end{figure}

In this case, the second SIT soliton is more intense and has a shorter temporal duration than the preparatory pulse, resulting in quite different behavior compared to the previous two cases. The second SIT soliton is still trapped, however, it generates a fast moving, high amplitude, transitory SIT soliton that is trapped much deeper in the medium than the location of the original excited region (see Fig.~\ref{fig:PrBigger}). Furthermore, the intensity of the negative control pulse, generated during the trapping of the second SIT soliton from the left hand side of the original coherence, is large enough to overwhelm the positive control light field, emitted from the right hand side of the original excited region. As a result, the transitory SIT soliton is captured by a negative control pulse and the phase of the ground state coherence remains constant between the two $\rho_{22}$ peaks.

Because the transitory SIT soliton has a higher intensity than the first SIT soliton, it appears as though the second SIT soliton travels through the excitation region associated with the first SIT soliton without any interaction, but, this is not the case. The second SIT soliton is trapped shallower in the medium than both the preceding cases, due to the strong control field amplification associated with the high peak intensity of the SIT soliton. However, the emitted control light field and transitory SIT soliton removes much of the excited population associated with the trapping, leaving the first $\rho_{22}$ region smaller than expected. As a result of these interactions, the second SIT soliton is trapped earlier, while the distance between the two excited regions is 14 times the FWHM of the original excitation region.

\subsection{Interaction between multiple trapped solitons and a control light field}
\label{ss:BiSolitons}
\begin{figure}
\begin{center}
\includegraphics[width=\columnwidth]{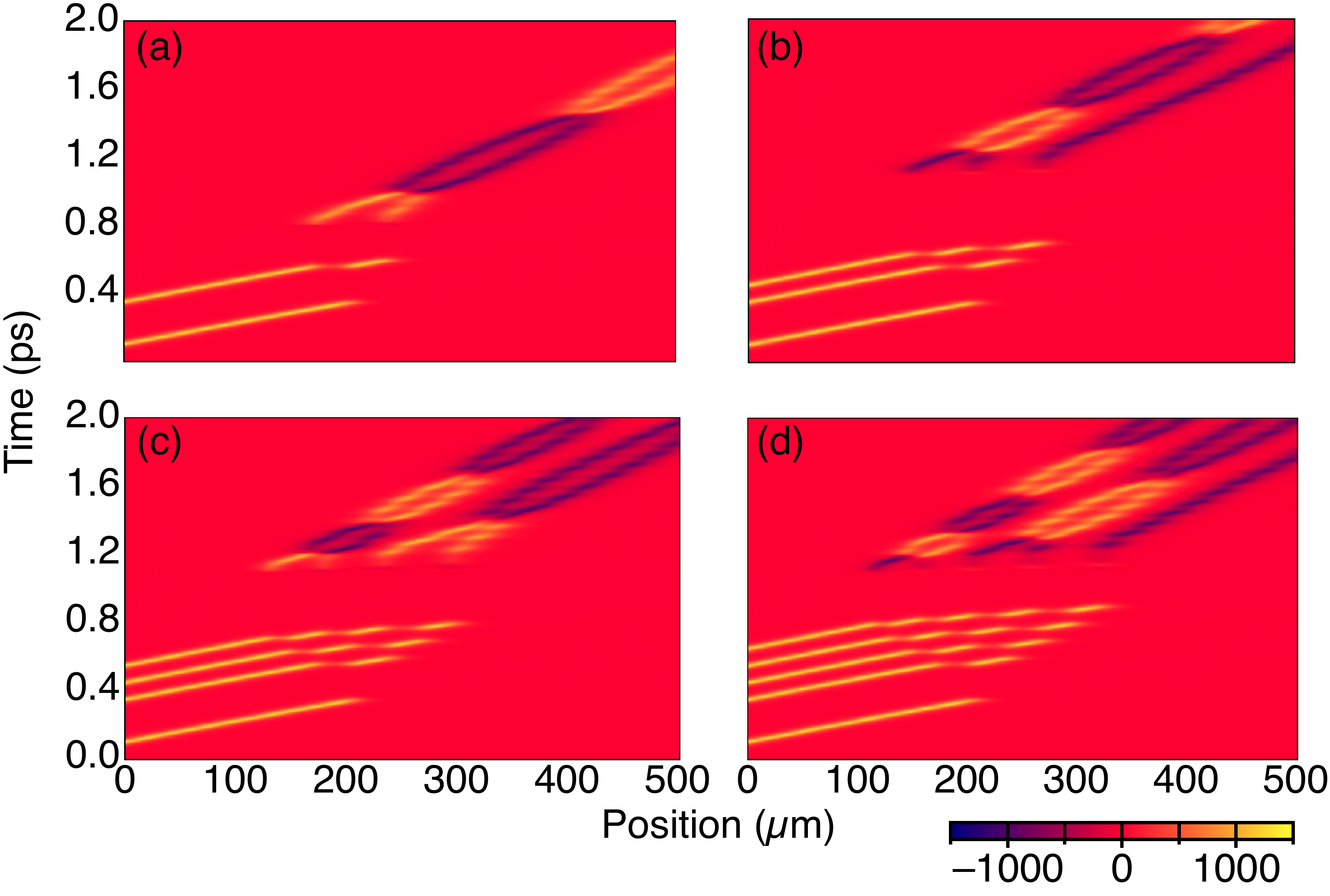}
\caption{\label{fig:SITsingleMolecule}The electric field $E_S$ (Vm$^{-1}$) throughout the medium in a set of calculations where two (a), three (b), four (c), and five (d) SIT solitons are trapped in the medium. Subsequently, from $t$=0.8~ns, a continuous wave control field is launched into the medium that interacts with the ground state coherence structure to form sets of bi-solitons and single solitons which propagate in the $+z$-direction. }
\end{center}
\end{figure}

As shown in Fig.~\ref{fig:pulseafterpi}, injecting a control light field alters the location of the captured soliton. In this section, we investigate this behavior after one SIT soliton has been captured by a control pulse and up to four subsequent SIT solitons have been trapped by the pre-existing ground state coherence. A control light field is then applied and the behavior of the trapped solitons is observed. In Fig.~\ref{fig:SITsingleMolecule}, the electric field, $E_S$, throughout the medium is shown for several different situations. 

Fig.~\ref{fig:SITsingleMolecule}(a) shows two SIT solitons propagating in response to the application of a cw control field after $\tau_0=0.8$~ns. The first SIT soliton is launched into the medium at $\tau_0=0.1$~ns and is captured by a small control pulse. At $\tau_0=0.4$~ns a second SIT soliton is launched into the medium, colliding with the ground state coherence $\rho_{12}$ structure (lower) as left over from the first SIT soliton and trapping it. From $t=0.8$~ns on a continuous wave control light field $E_c=0.5E_S$ is launched into the medium. This light field interacts with the ground state coherence structure as left by the two SIT solitons in the medium and generates a light field $E_S$ that propagates as a pair of coupled solitons in the positive $z$ direction. The ground state coherence (not shown in the figure) moves together with the pair as well.

More than two SIT solitons can be stopped and staggered in the medium, which is shown in Fig.~\ref{fig:SITsingleMolecule}, where (b) shows the case of three SIT solitons, (c) four and (d) five solitons stored. As can be seen, when the cw control field is injected from $t=0.8~ns$, only pairs of solitons are formed and that pairs do not cross to form larger structures. Furthermore, for an odd number of SIT pulses, one of the pulses (the last as seen from the entrance of the medium) does not pair up. 

This behavior can be understood as follows. The control field, which is absorbed by the first region of excited ground state coherence, is amplified by the second region and, as a result, has approximately the same amplitude after passing the excited regions (consisting of two trapped pulses) as before entering it. Furthermore, where multiple SIT pulses were trapped in the medium, each new pulse creates an excited region of ground state coherence that is shorter than the previous one. The control pulse interacts with these excited regions and creates a field $E_S$, where the power in the pulse is proportional to the total amount of ground state coherence with which the control pulse has interacted. The propagation speed of these created pulses is proportional to the power in the pulse. Therefore, the second soliton pair (as counted from the entrance of the medium) propagates faster than the first pair of solitons because it has a larger excited region and, as a result, generates more optical power. When an odd number of SIT pulses are inserted, the control field creates a single pulse in the $E_S$ field that propagates faster than the closest soliton pair.

Other systems (with different physical processes at work) in which such similar soliton structures propagate have been reported by Maruta and Hause \cite{maruta, hause} where these soliton structures were named bi-solitons. However, such bi-solitons in a $\Lambda$-type system have, to our knowledge, not been reported.

\subsection{Optical logic gates}
\label{ss:OpticalLogic}
Here, we show that an all-optical XOR gate can be implemented using combinations of SIT solitons and control pulses. Using the SIT soliton phase as the information carrier, an XOR operation is given by $\left|\phi_{1},\phi_{2}\right\rangle\rightarrow  \left|\phi_{1},\phi_{1}+\phi_{2}\right\rangle$. The logic operation begins with the cloning and capturing of the first SIT soliton ($\phi_{13}$), by the application of a control pulse ($\phi_{23}$). The phase difference between the control pulse and this SIT soliton is transferred to the ground state coherence of the medium at the cloning location (recall eq.~\ref{eq:eleven}), storing it. Thereafter, the SIT soliton with the phase to be operated on ($\phi_{2}$) is inserted into the medium. This SIT soliton collides with the stored coherence, generating a control pulse possessing a phase as described by eq.~(\ref{eq:twelve}). For the case that the first SIT pulse has a phase with the value zero, the resulting second control pulse that exits the medium has a phase which is the sum of the first small control pulse and the second SIT pulse as needed to perform the XOR operation.

\begin{eqnarray}
\theta=\phi_2+\phi_{23}-\phi_{13}
\label{eq:twelve}
\end{eqnarray}

Note that, in addition to encoding the result of the XOR operation in the phase of the emitted control pulse, the ground state coherence of the medium also stores the result of the operation. The stationary nature of the ground-state coherence then allows multiple operations to be performed. Furthermore, because each operation results in the medium emitting a pulse of light that encodes the result of that operation, conditional operations become feasible. 

It is interesting to estimate the number of optical logic operations that might be performed on a trapped SIT soliton. The limit to the number of operations that can be performed on a stored soliton can be estimated by comparing laser pulse repetition frequencies and ground state coherence times. An ultracold atomic ensemble quantum memory gate with a 1~ms decoherence time was recently demonstrated \cite{zhao}, while laser repetition rates exceeding 300~GHz have also been demonstrated \cite{merghem}. This provides an upper bound of 300 million operations on a stored SIT pulse.

\section{conclusions}
\label{s:conclusions}
We have analyzed the behavior of SIT solitons in a three-level medium by solving the density matrix and Maxwell equations. We have found that multiple SIT solitons can be stored as a ground state coherence in the three level medium by either the application of a control light field or by colliding the SIT soliton with a pre-existing ground-state coherence generated by a capture event. We have analyzed how the phase of the ground state coherence is related to the phase of the SIT soliton and the control pulse. The process of trapping SIT solitons has been investigated and has been shown to be the result of a complex process of repeated SIT soliton absorption and emission events. 

The propagation of trapped SIT solitons, driven by a control light field, has been investigated. As a result, bi-soliton behavior was observed for the first time with SIT solitons. Finally, we have shown that the phase of the SIT soliton can be used to encode information and demonstrate the operations required to provide an all-optical XOR gate. We note that the results from every operation on the ground-state coherence is stored in the ground-state coherence \emph{and} emitted as a control light field pulse.

\bibliographystyle{unsrt}
\end{document}